\documentclass[aps,prl,twocolumn,superscriptaddress,showkeys,showpacs]{revtex4}

\usepackage{amssymb}
\usepackage{amsfonts}
\usepackage{amsmath}
\usepackage{amsthm}
\usepackage{epsfig}
\usepackage{color}

\begin{document}

\title{Loss of coherence in dynamical networks: spatial chaos
and chimera states}

\author{Iryna Omelchenko}
\affiliation{Institut f{\"u}r Theoretische Physik, TU Berlin, Hardenbergstra\ss{}e 36, 10623 Berlin, Germany}
\affiliation{Institute of Mathematics, National Academy of Sciences of Ukraine, Tereshchenkivska str. 3, 01601 Kyiv,
Ukraine}
\author{Yuri Maistrenko}
\affiliation{Institute of Mathematics, National Academy of Sciences of Ukraine, Tereshchenkivska str. 3, 01601 Kyiv,
Ukraine}
\affiliation{National Center for Medical and Biotechnical Research, National Academy of Sciences of Ukraine,
Volodymyrska str.54, 01030 Kyiv, Ukraine}
\author{Philipp H{\"o}vel} 
\affiliation{Institut f{\"u}r Theoretische Physik, TU Berlin, Hardenbergstra\ss{}e 36, 10623 Berlin, Germany}
\author{Eckehard Sch{\"o}ll}
\affiliation{Institut f{\"u}r Theoretische Physik, TU Berlin, Hardenbergstra\ss{}e 36, 10623 Berlin, Germany}

\date{\today}
\preprint{OME11, draft from \today}
\begin{abstract}
We discuss the breakdown of spatial coherence in networks of coupled oscillators with nonlocal interaction. By
systematically analyzing the dependence of the spatio-temporal dynamics on the range and strength of coupling, we
uncover a dynamical bifurcation scenario for the coherence-incoherence transition which starts with the appearance of narrow layers of incoherence occupying
eventually the whole space.
Our findings for coupled chaotic and periodic maps as well as for time-continuous R{\"o}ssler systems reveal that 
intermediate, partially coherent states represent characteristic spatio-temporal patterns at the transition from
coherence to
 incoherence.\end{abstract}


\maketitle

Understanding the dynamics on networks is at the
heart of modern nonlinear science and has a wide applicability to
various fields~\cite{WAT98,SON10a}. Thus, network science is a vibrant,
interdisciplinary research area with strong connections
to physics. For example, concepts of theoretical physics
like the Turing instability, which is a known paradigm
of non-equilibrium self-organization in space-continuous
systems, have recently been transferred to complex
networks~\cite{NAK10}. While spatially extended systems show
pattern formation mediated by diffusion, i.e., local interactions,
a network takes also into account long-range
and global interactions yielding more realistic spatial
geo\-metries.

Network topologies like all-to-all
coupling of, for instance, phase oscillators  (Kuramoto model) or chaotic maps (Kaneko model) 
were intensively studied , and numerous characteristic regimes were
found~\cite{KUR84,MOS02,KAN96a}.  In particular, for globally coupled chaotic maps they range --for decreasing coupling
strength -- from complete chaotic synchronization
via clustering and  chaotic itineracy 
to complete desynchronization.
The opposite case, i.e., nearest-neighbor coupling, is known as lattice dynamical systems  of time-continuous
oscillators, or coupled map lattices if the oscillator dynamics is discrete in time. 
These kinds of networks arise naturally as 
discrete approximation of systems with diffusion and have also been thoroughly  studied. They can demonstrate rich
dynamics such as solitons, kinks, etc.~up to fully developed spatio-temporal chaos
\cite{JEN85,CHO95a,AFR05,NIZ02,KAN96a}.

The case of networks with nonlocal coupling, however, has been much less studied in spite of numerous applications in
different fields. Characteristic examples pertain to
neuroscience~\cite{BAT07,VIC08}, chemical oscillators~\cite{BET04a,MIK06}, electrochemical systems~\cite{Krischer},
and Josephson  junctions~\cite{WIE96a}. A new impulse to study
such networks was given, in particular, by the discovery
of so-called chimera states 
\cite{Chimera,OME10a}. 
The main peculiarity of these spatio-temporal
patterns is that they have a hybrid spatial structure, partially coherent and partially incoherent, which can develop
in networks of identical oscillators without any sign of inhomogeneity. 

In this Letter we discuss the transition between coherent and incoherent dynamics in networks of nonlocally coupled
oscillators. We start with coupled chaotic maps
\begin{equation}\label{eq:map}
z_i^{t+1} = f\left(z_i^t\right) + \dfrac{\sigma}{2P} \sum\limits_{j=i-P}^{i+P} \left[ f\left(z_j^t\right) -
f\left(z_i^t\right) \right],
\end{equation}
where $z_i$ are real dynamic variables ($i=1,...,N$,  $N \gg 1$ and the index $i$ is periodic mod $N$), 
$t$ denotes the discrete time, 
$\sigma$ is the coupling strength, $P$ specifies the number of neighbors in each direction 
coupled with the $i$-th element, and $f(z)$ is a local one-dimensional map.  
We choose  $f$ as the logistic map $f(z)=az(1-z)$  and  fix the bifurcation parameter $a$ at the value $a=3.8$. 
This choice yields chaotic behavior of the map $f$ with positive Lyapunov exponent $\lambda \approx 0.431$.

\begin{figure}[t]
\begin{center}
\includegraphics[width=0.9\linewidth]{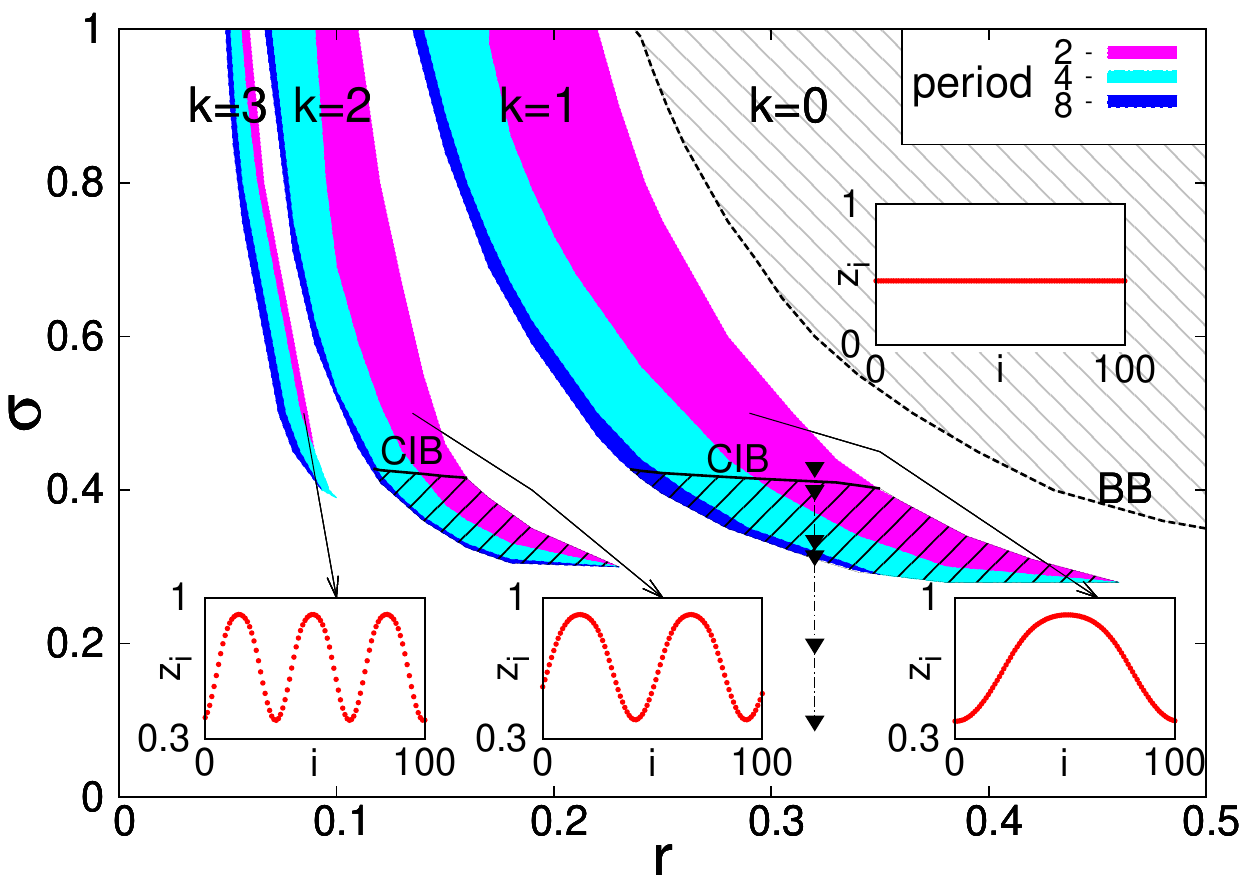}
\end{center}
\caption{(Color online) Regions of coherence for system~(\ref{eq:map}) in the $(r,\sigma)$ parameter plane with  wave
numbers $k=1, 2$, and $3$.  Snapshots of typical coherent states $z_i$ are shown in the insets.
Color code inside the regions distinguishes different time-periods of the states. 
The CIB bifurcation curve separates regions with coherent and incoherent dynamics. In the hatched and shaded (color)
regions below CIB two-cluster incoherent states exist. Completely synchronized chaotic states exist in the light hatched
region bounded by the blowout bifurcation curve BB. Parameters: $a=3.8$ and $N=100$.}
\label{Fig1}
\end{figure}

Results of direct numerical simulation of the model~(\ref{eq:map}) in the two-parameter plane of the coupling radius
$r=P/N$  and coupling strength $\sigma$ are presented in Fig.~\ref{Fig1}. This figure
reveals the appearance of regions of spatial coherence, shown in shading (color),
at an intermediate radius of coupling.
Alternatively, if the oscillators are uncoupled ($r=0$) or coupling is only local ($r = 1/N$) the network 
displays high-dimensional space-time chaos~\cite{KAN96a}.
In the opposite situation, when the coupling is all-to-all ($r=0.5$), chaotic synchronization occurs: the oscillators
behave identically, but chaotically in time following the dynamics of $f$ 
\cite{KAN96a,MOS02}. The chaotic synchronization (hatched region $k=0$) persists for
smaller $r$ or $\sigma$ up to the blowout bifurcation \cite{OTT94} indicated by the curve BB, where the synchronized
state loses transverse stability, i.e., the dynamics becomes desynchronized.
The spatially homogeneous state represents the simplest example of coherent dynamics.

In general we call a network state $z_i^t,$ $i=1,...,N$, \textit{coherent} on the ring $\mathcal{S}^1$ as $N
\rightarrow \infty$  if for any point $x\in\mathcal{S}^1$ the limiting value 
\begin{equation}
\lim\limits_{N \rightarrow \infty} \lim\limits_{t \rightarrow \infty} \sup\limits_{i,j \in U_{\delta}^N (x)} \left|
z_i^t - z_j^t \right|\rightarrow 0, \quad \text{for} \quad \delta\rightarrow 0,
\label{Eq:CoherenceDef}
\end{equation}
where $U_{\delta}^N (x) = \left\{ j:~0 \leq j \leq N,~ \left| {j}/{N} -x
\right| < \delta \right\}$ denotes a network-neighborhood of the point $x.$  If the limit~(\ref{Eq:CoherenceDef}) does
not vanish for $\delta\rightarrow 0$, at least for one point $x$, the network state is considered incoherent.  

Geometrically, coherence  means that in the thermodynamic limit $N\rightarrow \infty$ snapshots of the state $z_i^t$
 approach a smooth profile  $z(x,t)$ of the spatially continuous version of Eq.~(\ref{eq:map}) given by
\begin{equation}\label{eq:continuous}
z^{t+1}(x)= f(z^t(x))+ \dfrac{\sigma}{2r} \int_{x-r}^{x+r} \left[f\left(z^{t}(y)\right) -f\left(z^{t}(x)\right)\right] dy.
\end{equation}
According to the definition given above,  a transition from coherence to incoherence occurs when the respective solution
profile $z^t(x)$ 
becomes discontinuous.  
Note that for networks of phase oscillators  the property of coherence and incoherence can also be established
with use of  the notion of a local order parameter~\cite{PIK08,WOL11a}.

Regions of coherence and  typical shapes of the respective coherent states $z_i^t$ are shown in Fig.~\ref{Fig1} as
shaded (color) tongues and insets, respectively. A coherent state has a smooth profile characterized by the number of
maxima,
i.e., the wave number $k$. Only regions for wave numbers $k = 1, 2,$ and $3$ are shown.  Further decrease of $r$
yields additional thin higher-order regions following a period-adding cascade $k =4,5,..$. 
Inside the regions, the states are coherent in space and periodic in time and undergo a period-doubling cascade of 
bifurcations in time as $r$ or $\sigma$ decrease.  
In the parameter space between the coherence regions  the
network dynamics remain  coherent but not periodic anymore.  
The states alternate chaotically between the adjacent $k$-states and thus exhibit chaotic
itineracy~\cite{KAN96a,KAN03}. The combination of  {\it period-adding} in space and {\it period-doubling} in time
represents a remarkable feature of networks of coupled chaotic oscillators with nonlocal coupling.

\begin{figure}[t]
\includegraphics[width=\linewidth]{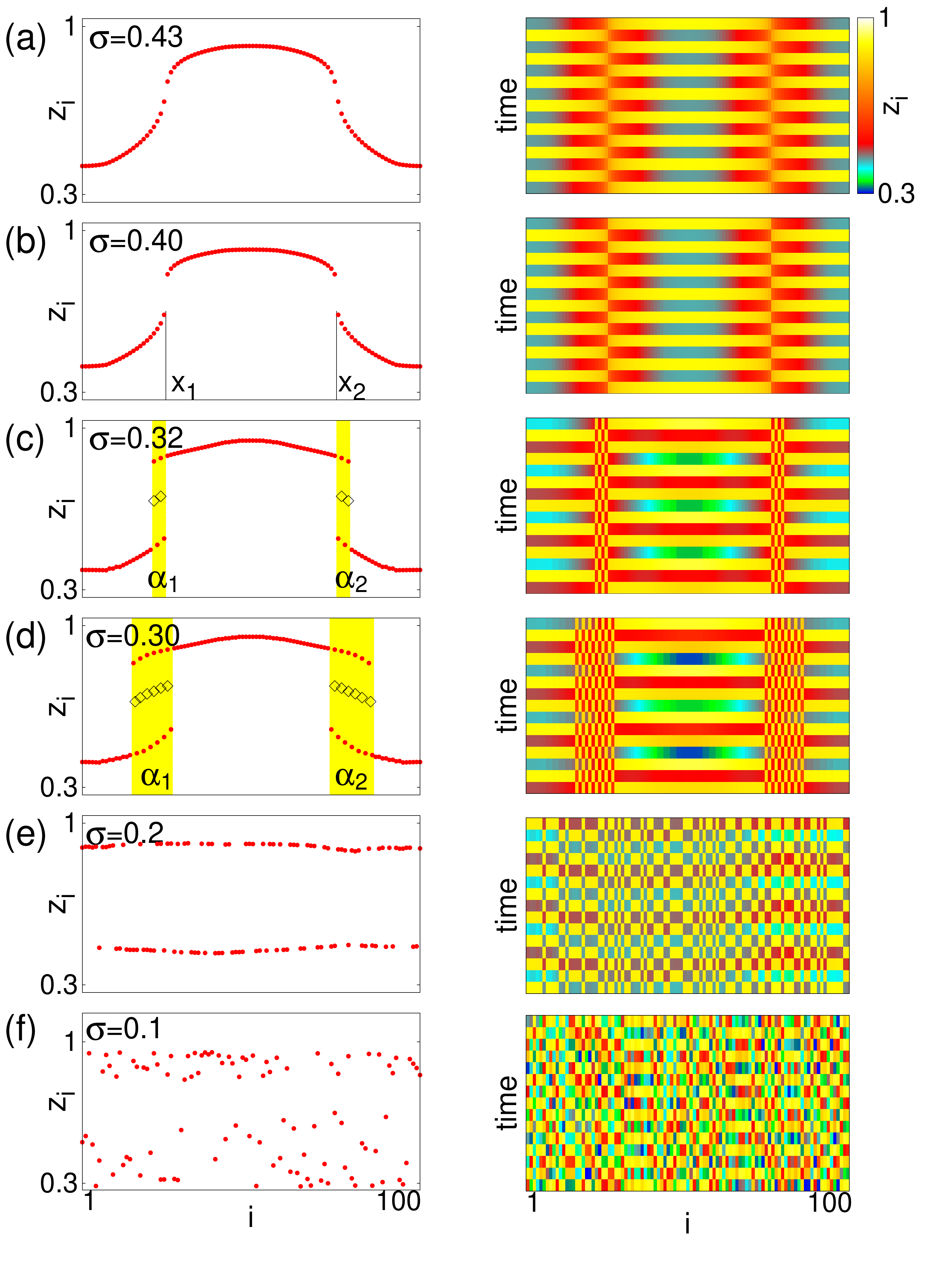}
\caption{(Color online) Coherence-incoherence bifurcation for coupled chaotic logistic maps
for fixed coupling radius $r=0.32$ (black triangles in Fig.~\ref{Fig1}). 
For each value of the coupling parameter $\sigma$ 
(decreasing from top to bottom, $\sigma= 0.43, 0.4, 0.32, 0.3, 0.2,$ and $0.1$, respectively)
snapshots (left columns) and space-time plots (right columns) are shown.
Other parameters as in Fig.~\ref{Fig1}.}
\label{Fig2}
\end{figure}

A typical scenario of the coherence-incoherence transition is illustrated in Fig.~\ref{Fig2}(a)-(f),  where we fix the
coupling radius $r=0.32$ and decrease the coupling strength $\sigma$ along the vertical line with triangles in
Fig.~\ref{Fig1}. First, in Fig.~\ref{Fig2}(a), the solution profile $z_i^t$ is clearly smooth for $\sigma=0.43$. Thus,
the network dynamics is spatially coherent. 
For smaller $\sigma$, the profile $z_i^t$ sharpens up and, at some value $\sigma \cong 0.40$,  
loses smoothness in two points  $x_1$  and $x_2$ as shown in Fig.~\ref{Fig2}(b).
This is a bifurcation point for the coherence-incoherence transition: Beyond this parameter value, the wave-like profile
$z_i^t$ splits up into upper and lower branches, and two narrow boundary layers of incoherence are born around the
points $x_1$ and $x_2$ (shaded yellow stripes  $\alpha_1$ and $\alpha_2$ in Fig.~\ref{Fig2}(c)). 
The incoherence stripes become wider with further decrease of $\sigma$ (Fig.~\ref{Fig2}(d)) and, eventually, the
dynamics
becomes completely incoherent (Figs.~\ref{Fig2}(e) and (f)).

\begin{figure}[t]
\includegraphics[width= \linewidth]{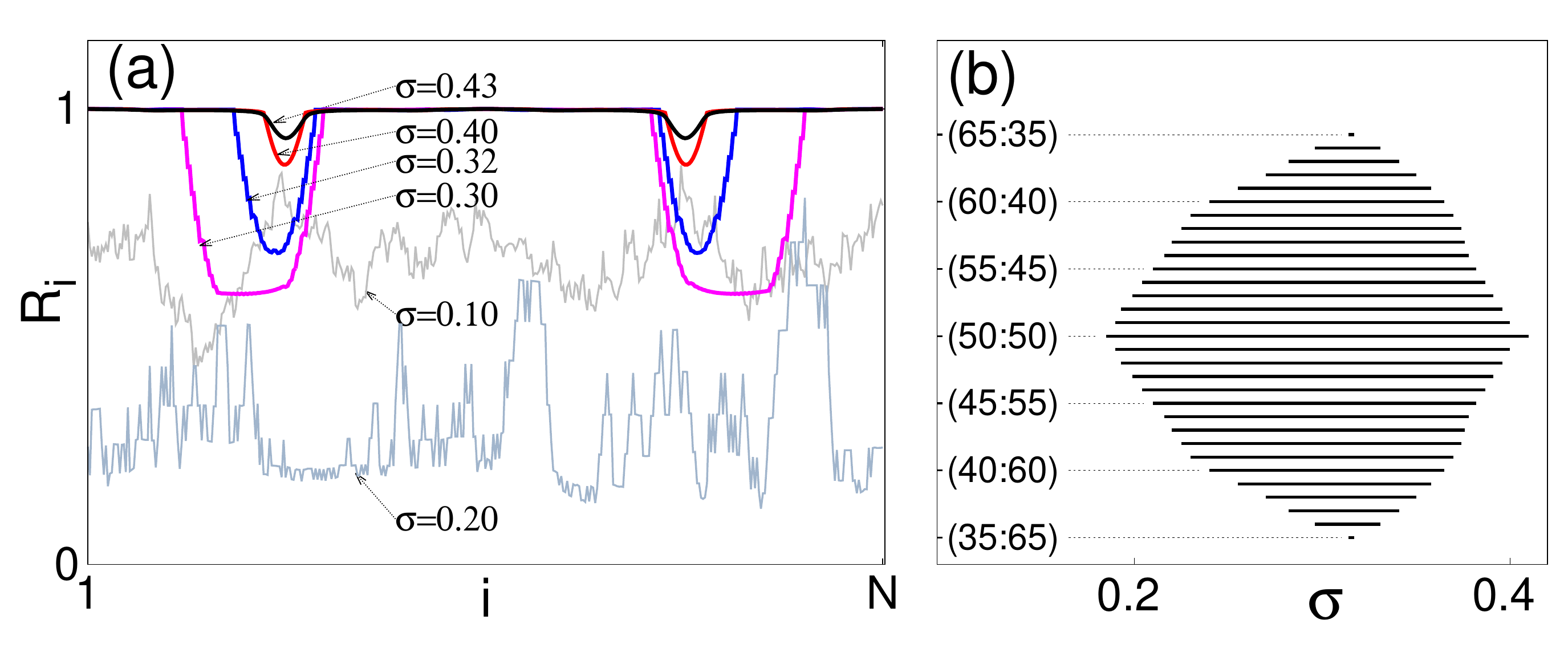}
\caption{(Color online) (a): Local order parameter for $r=0.32$ and different $\sigma$ as in Fig.~\ref{Fig2}
(approximated with $\delta=10, N=400$).  (b): Regions of stability for the two-cluster solutions with different 
asymmetries $(n_1:n_2)$ as a function of $\sigma$ for $r=0.35$. 
Other parameters as in Fig.~\ref{Fig1}.}
\label{Fig3}
\end{figure}

In our numerical simulations, no coherent states were found below the bifurcation parameter value  $\sigma \cong
0.40$. In contrast, numerous hybrid states arise, which are coherent on some intervals of the ring $\mathcal{S}^1$ 
and incoherent on the complementary intervals. Typical examples of these partially coherent states are shown in 
Figs.~\ref{Fig2}(c) and (d). In the figure, black diamonds mark a threshold within the incoherent
regions:  If the initial value for a chosen oscillator is located above/below this diamond, with all other
oscillators unchanged, it will be attracted by the upper/lower branch. This implies that within the
incoherent intervals
$\alpha_1$ and $\alpha_2$ any combinations of the upper and lower states-- so-called \textit{mosaic} \cite{CHO95a}
or \textit{skeleton} pattern \cite{NIZ02} --  are  admissible and can be realized by appropriate choice of
the initial
conditions. The coherence-incoherence transition is illustrated by the local order parameter $R_i$ shown in 
Fig.~\ref{Fig3}(a) for the snapshots depicted in Fig.~\ref{Fig2}. It is defined as (cf. Ref.~\cite{WOL11a})
\begin{equation}
 R_i=\lim_{N\rightarrow\infty}\frac{1}{2\delta(N)}\left|\sum_{|j-i|\leq\delta}e^{i\psi_j}\right|, ~~~ (i=1,\dots,N)
\end{equation}
with the phase $\psi_j$ introduced by the mapping $\sin\psi_j=(2z_j-\max_jz_j-\min_jz_j)/(\max_jz_j-\min_jz_j)$
and $\delta/N\rightarrow0$ for $N\rightarrow\infty$, such that a spatial half-period oscillation is mapped onto the
 polar angular interval $[-\pi/2, \pi/2]$. $R_i$ is close to unity for the coherent state and decreases in
regions of spatial incoherence.
For the two cases of complete incoherence (Fig.~\ref{Fig2}(e),(f)) the local order parameter is much
smaller than unity, and fluctuating strongly as a signature of spatial chaos. In case of very small coupling
(Fig.~\ref{Fig2}(f)) the values of $z_i$ are more spread out, and hence $R_i$ varies less strongly between neighboring
sites and is on average larger than in
Fig.~\ref{Fig2}(e).

As it is illustrated in the space-time plots of Fig.~\ref{Fig2}, the  temporal dynamics before and after the
coherence-incoherence bifurcation remains periodic up to very small coupling, when it is chaotic (Fig.~\ref{Fig2}(f)).
The system's complexity results from the fact that the bifurcation
gives rise to a huge multistability of partially coherent states as $N\rightarrow\infty$. Indeed, the number $c_N$ of
different partially coherent states born in the  bifurcation is $c_N=2^{d N}$, where $d$ is the fraction of
oscillators in 
the incoherent part ($d=\alpha_1 + \alpha_2$ in the case of two incoherence intervals as in Figs.~\ref{Fig2}(c) and
(d)). It follows that the number of different states grows exponentially fast  with $N$,  and  the \textit{spatial entropy}
$h$,
which is defined as  $h = \lim_{N \rightarrow \infty} (1/N) \ln c_N$, is positive and equals $h=d
\ln 2$. Positive spatial entropy means that the system displays {\it spatial chaos}~\cite{CHO95a,AFR05,NIZ02}, i.e.,
sensitive dependence
on space coordinates.
Therefore, the coherence-incoherence bifurcation results instantly in the appearance of spatial chaos that develops
first at narrow incoherence intervals and, with decreasing $\sigma$, spreads
onto the whole ring. Thus a chimera-like state of coexisting coherent and incoherent regions arises as a transitional
state in the coherence-incoherence bifurcation scenario. However, in contrast to previously reported chimera states in
time-continuous systems \cite{Chimera,OME10a}, the temporal behavior is periodic rather than chaotic, and the complexity
arises
due to the huge variety of multistable incoherent states corresponding to permutations of the sequence of upper and
lower local states. With further decrease of $\sigma$, the chimera states disappear giving rise to completely
incoherent
behavior.

To identify the parameter range for partially coherent states, 
we define a mosaic of $z_i^t$
as a symbolic sequence  of
"$-$" or "$+$" if the value $z_i$ belongs 
to the lower or upper branch of the solution profile, respectively.  
Hence, states as in Fig.~\ref{Fig2}(b) are given by the mosaic of the form $(...--++...++--...)$.
Therefore, they may be considered as two-cluster states with the ratio of $(n_1:n_2)$,  where $n_1$ and $n_2$ are the
numbers of "$-$" and "$+$" ($n_1 + n_2=N$), respectively. The ratio $(n_1:n_2)$ indicates the level of asymmetry of the
solution $z_i^t$,  which is important for its stability. Indeed, as it is illustrated in Fig.~\ref{Fig3}(b) for
$r=0.35$, 
the symmetric solution $(50:50)$ has the widest stability interval 
$\sigma \in (0.189,0.41)$.  As the asymmetry grows, the stability interval
shrinks, and the solution with the mosaic ratio $(35:65)$ has the shortest 
stability interval $\sigma \in (0.314,0.317)$. Two-cluster solutions with larger asymmetry cannot 
be stabilized anymore. The states with more complex mosaics, examples of which are presented in Fig.~\ref{Fig2}(c)-(e),
are characterized by more involved mechanisms of stability.

\begin{figure}[t]
\includegraphics[width=0.95\linewidth]{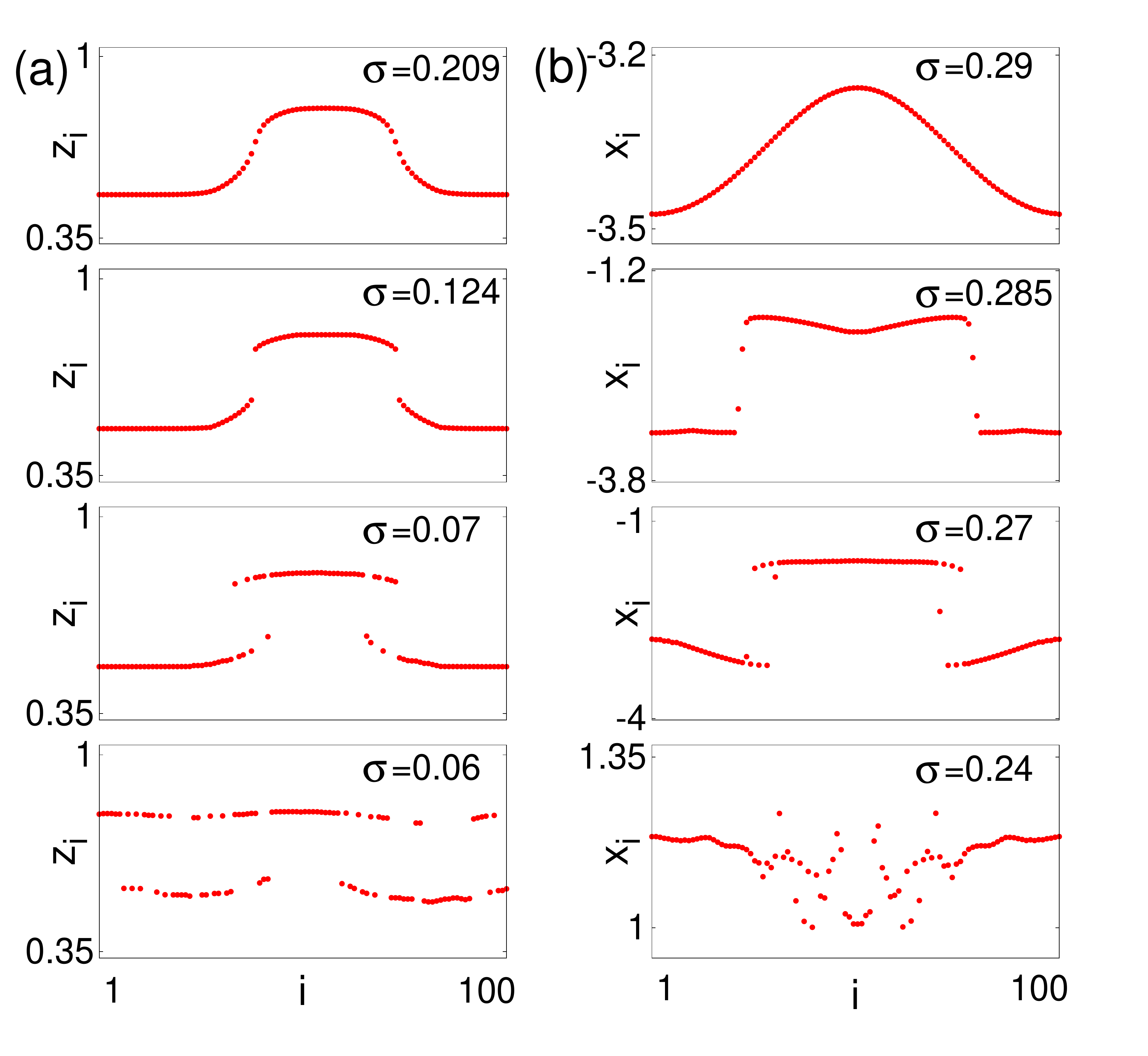}
\caption{(Color online) Snapshots with decreasing coupling strength $\sigma$: Coherence-incoherence transition for a ring of (a)   100 coupled periodic logistic maps ($a=3.2$) for a
coupling radius $r=0.1$ and (b) 100  nonlocally coupled R{\"o}ssler systems ($a=0.42$, $b=2$, $c=4$) with 
$r=0.3$.}
\label{Fig4}
\end{figure}

To test if the coherence-incoherence bifurcation is a universal scenario, we have also investigated
nonlocally coupled networks with different local dynamics.
Figure~\ref{Fig4} shows the coherence-incoherence transition for nonlocally coupled 
logistic maps (\ref{eq:map}) in the periodic regime ($a=3.2$, period $2$, Fig.~\ref{Fig4}(a))
 and for nonlocally coupled chaotic R{\"o}ssler
systems 
\begin{equation}\label{eq:roessler}
\begin{array}{l}
\dot{x}_i = -y_i-z_i + \dfrac{\sigma}{2P} \sum\limits_{j=i-P}^{i+P} \left( x_j - x_i \right), \\
\dot{y}_i = x_i + ay_i, \\
\dot{z}_i = b+z_i(x_i-c) \qquad \qquad \qquad (i =1,...,N)
\end{array}
\end{equation}
(Fig.~\ref{Fig4}(b)). As it can be seen from the snapshots, both models display a transition from spatial coherence to
incoherence, as the coupling strength $\sigma$ decreases, according to the scenario described above.
Since network~(\ref{eq:roessler}) is time-continuous, the oscillators within the incoherence intervals are not only
located at an upper or lower branch of the solution profile, but vary continuously.
This gives rise to chaotic temporal dynamics in the incoherent intervals, which resembles known
chimera states
\cite{Chimera,OME10a}.
We conclude that chaotic chimera states typically arise in the nonlocally coupled R{\"o}ssler systems, similar to
nonlocally coupled Kuramoto-Sakaguchi phase oscillators \cite{OME10a}.

In conclusion, we have identified a novel mechanism for the coherence-incoherence transition 
in networks with nonlocal coupling of variable range.  It consists in the appearance of multistable chimera-like
states. We have found similar  bifurcation scenarios for coupled maps with 
both chaotic and periodic local dynamics as well as for time-continuous systems. This indicates
a common, probably universal phenomenon in networks of very different nature, due to nonlocal coupling.

\begin{acknowledgments}
We thank B.~Fiedler, M.~Hasler, and M.~Wolfrum for illuminating discussions.
I.~O. acknowledges support from DAAD and DFG (SFB~910). 
Y.~M. acknowledges support and hospitality from TU Berlin.
\end{acknowledgments}


\begin{thebibliography}{10}

\bibitem{WAT98}
D.~J. Watts and S.~H. Strogatz, Nature {\bf 393},  440  (1998).

\bibitem{SON10a}
C. Song, T. Koren, P. Wang, and A.-L. Barab\'asi, Nature Physics {\bf 6},  818
  (2010).

\bibitem{NAK10}
H. Nakao and A.~S. Mikhailov, Nature Physics {\bf 6},  544  (2010).

\bibitem{KUR84}
Y. Kuramoto, {\em Chemical Oscillations, Waves, and Turbulence} (Springer, Berlin Heidelberg, 1984), 
J.A. Acebr{\'o}n et al. Rev. Mod. Phys. {\bf 77}, 137 (2005).

\bibitem{MOS02}
E. Mosekilde, Y. Maistrenko, and D. Postnov, {\em Chaotic Synchronization:
  Applications to Living Systems} (World Scientific, Singapore, 2002).

\bibitem{KAN96a}
K. Kaneko and I. Tsuda, {\em Chaos and Beyond, A Constructive Approach with
  Applications in Life Sciences} (Springer, Berlin, 1996)

\bibitem{JEN85}
M. H. Jensen, Physica Scripta {\bf T9}, 64 (1985); 
P. Coullet, C. Elphick, and D.~Repaux, Phys. Rev.
Lett. {\bf 58}, 431 (1987); W. Shen, SIAM J.~Appl. Math. {\bf 56}, 1379 (1996); B.~Fernandez, B.~Luna, and E.~Ugalde,
Phys. Rev. E {\bf 80}, 025203(R) (2009).

\bibitem{CHO95a}
S.~N. Chow and J. Mallet-Paret, IEEE~Trans.~Circ.~Sys. {\bf 42},  746  (1995).

\bibitem{AFR05}
V. Afraimovich, {\em Some topological properties of lattice dynamical systems} in Dynamics of coupled map lattices and of related
spatially extended systems, Lecture Notes in Phys. {\bf 671}, 153 (Springer, Berlin, 2005).

\bibitem{NIZ02}
L.~P. Nizhnik, I.~L. Nizhnik, and M. Hasler, Int.~J.~Bif.~Chaos {\bf 12},  261
  (2002).

\bibitem{BAT07}
D. Battaglia, N. Brunel, and D. Hansel, Phys.~Rev.~Lett. {\bf 99},  238106
  (2007).

\bibitem{VIC08}
R. Vicente, L.~L. Gollo, C.~R. Mirasso, I. Fischer, and P. Gordon, Proc. Natl.
  Acad. Sci. {\bf 105},  17157  (2008).

\bibitem{BET04a}
C. Beta, M.~G. Moula, A.~S. Mikhailov, H.~H. Rotermund, and G. Ertl, Phys. Rev.
  Lett. {\bf 93},  188302  (2004).

\bibitem{MIK06}
A.~S. Mikhailov and K. Showalter, Phys.~Rep. {\bf 425},  79  (2006).

\bibitem{Krischer}
N. Mazouz, G. Fl{\"a}tgen, and K. Krischer, Phys.~Rev.~E {\bf 55},  2260
  (1997); V. Garc\'{\i}a-Morales and K. Krischer, Phys. Rev. Lett. {\bf 100},  054101
  (2008).

\bibitem{WIE96a}
K. Wiesenfeld, P. Colet, and S.~H. Strogatz, Phys. Rev. Lett. {\bf 76},  404
  (1996).

\bibitem{Chimera}
Y. Kuramoto and D. Battogtokh, Nonlin. Phen. in Complex Sys. {\bf 5},  380
(2002); D.~M. Abrams and S.~H. Strogatz, Phys.~Rev.~Lett. {\bf 93},  174102  (2004);
G.~C. Sethia, A. Sen, and F.~M. Atay, Phys.~Rev.~Lett. {\bf 100},  144102
(2008); C.~R. Laing, Physica D {\bf 238},  1569  (2009); E.~A. Martens, C.~R. Laing, and S.~H. Strogatz, Phys.
Rev. Lett. {\bf 104}, 044101 (2010).

\bibitem{OME10a}
O.~E. Omel'chenko, M. Wolfrum, and Y.~L. Maistrenko, Phys. Rev.~E {\bf 81},
  065201  (2010).

\bibitem{OTT94}
E. Ott and J. C. Sommerer, Phys. Lett. A \textbf{188}, 39 (1994). 

\bibitem{PIK08}
A.~S. Pikovsky and M.~G. Rosenblum, Phys. Rev. Lett. {\bf 101},  264103 (2008).

\bibitem{WOL11a}
M. Wolfrum, O.~E. Omel'chenko, S. Yanchuk, and Y.~L. Maistrenko, Chaos \textbf{21}, 013112 (2011).

\bibitem{KAN03}
K. Kaneko and I. Tsuda, Chaos {\bf 13}, 926 (2003).

\end{thebibliography}
\end{document}